# Structure and Needed Properties of Reasonable Polarization Mode Dispersion Emulators for Coherent Optical Fiber Transmission

Reinhold Noé[(1,2)], Benjamin Koch[(1,2)]
(1) Paderborn University, EIM-E, Warburger Str. 100, D-33098 Paderborn, Germany, e-mail: noe@upb.de
(2) Novoptel GmbH, Helmerner Weg 2, D-33100 Paderborn, Germany, E-mail: info@novoptel.com

*Abstract*—This paper proposes a scientifically reasonable polarization mode dispersion (PMD) emulator (PMDE) for coherent optical fiber transmission.

Guidelines are physically correct modeling of the polarization-dispersive fiber, the time-variable polarization transformations occurring in there, including emulation of polarization events caused by lightning strikes, the adoption of acceptable compromise to keep implementation cost low enough and competitive industrial basis for production of such PMDE.

We propse a PMDE consisting of $N$ differential group delay (DGD) sections placed between $N+1$ time-variable general retarders or polarization scramblers. These should be general elliptical retarders, capable of changing polarization with rates up to 20 Mrad/s on the Poincaré sphere. That should include bursts of polarization rotations forth and back at up to 20 Mrad/s. The DGD sections can be fixed or variable and should be able to constitute a total PMD of alternatively, say, 20 ps, 50 ps, 100 ps, 200 ps, or another set of various discrete values. We propose even $N$ and equal individual DGDs, which allows the total PMDE to assume a neutral state without any PMD of whatever order. Number $N$ may be chosen relatively small; $N = 2$ seems acceptable. A variety of component and subsystem suppliers is available, and the proposed PMDE is available on the market.

## I. Introduction

The dominance of coherent optical polarization division multiplex transmission on amplified fiber lines, usually with non-negligible polarization mode dispersion (PMD) [1], and the observed occurrence of fast polarization transformation changes, namely due to lightning strikes, make it advisable or necessary to standardize polarization mode dispersion emulators (PMDE). We review available knowledge in this area and propose structure as well as properties to be implemented for such PMDEs.

## II. Modeling of PMD

The existence of first-order polarization mode dispersion was laid out in 1986 by Poole and Wagner [1]. They found that there exist two principal states of polarization (PSP) with maximum and minimum group delay, hence a differential group delay (DGD). They derived this by showing that, upon application of a PSP at the fiber input, the output polarization does, to first order, not vary as a function of frequency. PSPs were given as Jones vectors.

If the PSPs are constant along the fiber which accumulates DGD then it may be called a DGD section. Polarization-maintaining fibers (PMF) or other highly birefringent devices such as Z-cut $LiNbO_3$ waveguides are (quasi-)ideal DGD sections.

In 1999 one of us showed that the small-signal intensity modulation transfer function yields minimum and maximum group delays when the PSPs are launched [2], and PMD can be entirely derived from there. The PSPs were expressed as normalized Stokes vectors. This makes sense because by that time it had been recognized in the scientific community that total first-order PMD of 2 DGD sections can be calculated by the law of cosines, i.e. through a vectorial addition of the two individual PMD vectors. We pointed out [2] (in a slightly different nomenclature) that this can be generalized to an arbitrary number of DGD sections,

$$\tilde{\Omega} = \sum_{i=1}^{N} \tilde{\Omega}_i \qquad \tilde{\Omega}_i = \left(\prod_{j<i} \mathbf{G}_j^T\right)\mathbf{\Omega}_i . \qquad (1)$$

Here $\tilde{\Omega}$ is the overall, total input-referred PMD vector, $\tilde{\Omega}_i$ are the individual input-referred PMD vectors of the various DGD sections, and $\mathbf{G}_j$ (with $^T$ = transpose) are the rotation matrices of all retarders preceding DGD section $i$, including preceding DGD sections. In each case, the (total or individual) DGD equals the length of a PMD vector ($\tilde{\Omega}$ or $\tilde{\Omega}_i$). A proof of the PMD vector concatenation rule is laid out in [3 (p. 96)].

The $\tilde{\Omega}_i$ can be graphically concatenated in a PMD or DGD profile [2] which visualizes PMD in the normalized Stokes space that is scaled in temporal DGD units (typically in ps). An exemplary DGD profile is shown in Fig. 1 top.

In 1991, Foschini and Poole [4] have expanded the overall PMD vector by a Taylor series in the frequency domain,

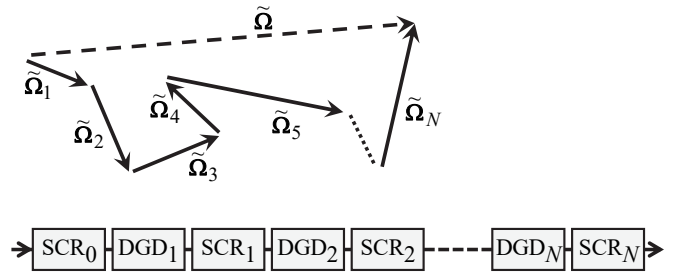

Fig 1. Top: Exemplary differential group delay profile, with overall first-order PMD being equal to sum of individual DGD or PMD vectors. Bottom: Proposed PMD emulator with $N$ DGD sections (indices 1...$N$) placed between $N+1$ retarders/scramblers SCR (indices 0...$N$).

$$\widetilde{\Omega}(\omega) = \widetilde{\Omega}(\omega_0) + (\omega - \omega_0)\widetilde{\Omega}'(\omega_0) + \frac{(\omega - \omega_0)^2}{2!}\widetilde{\Omega}''(\omega_0) + \ldots \quad (2)$$

This eqn. (2) has dominated the discussion of higher-order PMD ever since, with probably >100 papers written about it. However, the truncated Taylor expansion of the PMD vector is unphysical: Far off the considered optical angular frequency $\omega_0$ (which usually is the carrier frequency) it inevitably predicts $\left|\widetilde{\Omega}(\omega)\right| \to \infty$, hence an infinite overall DGD. This is incompatible with the reasonable assumption that at all angular frequencies $\omega$ the expected DGD as well as the maximum DGD are similar or even identical (though this occurs at different times or for different fiber temperatures etc.).

The DGD sections and retarders which physically model the transmission line create a **quasi-periodic** trajectory of the total PMD vector in the Stokes space (that is scaled in DGD units). It is well known that a Taylor series is a very bad choice for the modeling of a periodic or quasi-periodic function!

An alternative to the unphysical representation of PMD by (2) is the usage of a finite number of DGD sections. PMD compensators have been built this way [5–7]. In [5, 6, 7] the number $N$ of DGD sections was 3, 32 and 123, respectively. The latter large $N$ suggests the usage of, or rather is, a DGD profile which is a rod that can be bent. In [8], PMD modeling by a DGD profile that can be bent has been numerically tested against the Taylor expansion (2), with the same number of degrees-of-freedom (DOF). Of course the bent DGD profile outperforms the Taylor expansion by far, in terms of modeling accuracy. A sequence of DGD sections is similar to the bent DGD profile but is much easier realized with off-the-shelf components. Also, with a sequence of sections whose DGDs can be changed one is not restricted to a fixed total DGD like in [6, 7].

For the standardization of a PMD emulator we strongly recommend a finite sequence of $N$ DGD sections, placed between $N+1$ retarders SCR (= polarization scramblers), see Fig. 1 bottom. The scrambler eigenmodes and retardations determine the angles between subsequent DGD sections.

### III. NUMBER OF RETARDERS AND DGD SECTIONS

In the old days of intensity modulation, no polarization transformer or retarder was needed between transmitter (TX) and receiver (RX) as long as there was no PMD.

For coherent transmission which is inherently polarization-sensitive, a retarder between TX and RX is needed to model the fiber, even in the absence of PMD. This retarder and all retarders between which DGD sections are placed should be time-variable.

For emulation (or compensation) of just first-order PMD, one DGD section and another time-variable retarder must be added before the RX. This is in agreement with early 1st-order PMD compensators (which needed no second retarder right before the RX because the RX was polarization-insensitive).

From Section II. it is clear that a large $N$ would be best. In practice, cost limits the permissible number $N+1$ of time-variable retarders. Fortunately it is also technically reasonable to work with a fairly limited number $N+1$ of retarders:

Firstly, strong/frequent/fast polarization fluctuations can sometimes be generated at specific places (such as bridges, exchange offices). This is taken into account by our model [2] with limited number $N$ of DGD sections. Other researchers have come to the same conclusion and have called this a hinge model of PMD [9–12].

Secondly, large PMD is more difficult to equalize in the coherent RX than small PMD. Furthermore, beyond a certain amount of DGD the coherent RX will not be able to follow/track/equalize anyway. So, a Maxwellian DGD distribution with infinite tails is not needed. Rather it suffices to find that amount of PMD which is still tolerable. Then one knows that more than this cannot be tolerated, no matter how frequently or rarely that occurs. Since a DGD distribution tail of infinite length is not needed, a finite total DGD suffices. Such can be generated with a limited number $N$ of DGD sections. Of course higher order PMD effects get weaker when the DGD sections must form a larger overall DGD, but the same holds (though usually to less degree) also in the fiber whose PMD is emulated.

Already $N = 2$ cascaded DGD sections require an infinite Taylor series for exact modeling but as mentioned above the Taylor expansion of the PMD vector is unphysical anyway. The important thing is that 2 or more cascaded DGD sections generate PMD which is not limited to 1st order.

The larger $N$, the better. But already $N = 2$ (Fig. 2 top) seems to be a good choice, also due to cost reasons.

One may consider to use the PMDE in both directions, thereby effectively doubling available 1st order PMD. This would require 2 circulators (Fig. 2 bottom). However, since the available number of DOF does not increase the result is not expected to be as good as that which one could expect from a PMDE with $2N$ sections.

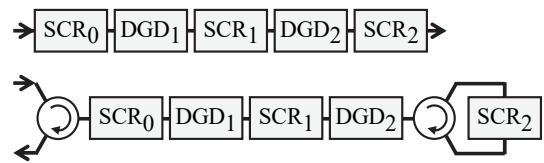

Fig. 2. Top: PMDE with $N = 2$ sections. Bottom: Larger PMDE with circulators and forth/back transmission.

### IV. TYPE AND SPEED OF TIME-VARIABLE RETARDERS

Each time-variable retarder is a polarization scrambler. A general elliptical retarder has 3 DOF, for example azimuth angle and elevation angle of one eigenmode and the retardation. According to [2], only one retarder in a PMDE must be an elliptical retarder, for instance the last one (index $N$). In that case it suffices for the others (indices $0...N–1$) that they be able to endlessly transform any input polarization into a PSP of the following DGD section. While such retarders need only 2 DOF it is usually too complicated to implement them with just the right 2 DOF. This holds all the more since between retarder and subsequent DGD section there can, in praxis, be an unknown polarization transformation. Hence we recommend that all time-variable retarders/scramblers of the PMDE preferably be general elliptical retarders with 3 DOF and endless polarization transformation properties.

To generate the recommended needed 3 DOF, more DOF are generally needed in practice. 4 or more rotating waveplates, such as quarterwave plates (QWP), are likely to be an adequate choice.

It is also possible to use fiber squeezers or equivalent technologies of retarders with fixed eigenmodes. However, it is difficult to make them fulfill the requirement that they be able to endlessly transform any input polarization into a PSP of the following DGD section. In particular, for reliable operation this usually requires fiber squeezers with calibrated retardation (i.e. temperature- and ageing-insensitive offset) [13], which is easily achieved if the actuators are magnets, but not if the actuators are piezos.

More in detail, to prove endless polarization transformation capability of a polarization transformer one must be able to operate it as an endless controller/tracker with quantifiable low outage probability at the highest possible speed. Examples are given in [14]: Relative intensity error 0.45% is surpassed with probability of $10^{-10}$ when tracking 40 krad/s. Relative intensity error 1.2% is surpassed with probability of $10^{-12}$ when tracking 100 krad/s. Note that this does not preclude endless polarization scrambling up to very much higher speeds, typically 2...3 orders of magnitude higher, mainly because the delay by the accurate search algorithm and feedback is avoided.

In [15], fiber polarization fluctuations due to lightning strikes with speeds up to 5.1 Mrad/s have been observed. This study is very valuable for the estimation of the needed time-variance of the retarders in the PMDE. There is no proof that no stronger lightning strikes can occur, nor that they cannot take place at worse positions (i.e. nearer to the fiber).

The time-variable retarder should be able to replicate the polarization fluctuations caused by lightning strikes. Since they may be large, an endless polarization change is needed. That can be generated by phase shifters with infinite retardation, which unfortunately do not exist, or by rotating waveplates, namely a rotating halfwave plate (HWP). A rotating HWP behaves like a variable circular retarder cascaded with a HWP at fixed position. It happens that circular retardation indeed coincides with the Faraday effect in fiber, caused by the magnetic field that is generated by a lightning strike.

In order to cover the range of observed polarization changing speeds with some headroom, a scrambling speed of 20 Mrad/s seems adequate. If faster polarization changes in fiber should be observed this value may need to be revised upward. This would be possible, given that 50 Mrad/s polarization scramblers are on the market [16].

A fast rotating HWP which dominates the scrambling speed distribution can be beneficial [17]: There a polarization tracking system was tested roughly 50 times faster than with a Rayleigh-like speed distribution. This is considered as a tremendous advantage in the PMDE process, given that most likely only certain combinations of the $N+1$ scrambler speeds and positions will cause worst performance in coherent transmission links. Of course the various HWP and other waveplate rotation speeds should all be incommensurate.

In agreement with the above one retarder/scrambler waveplate should be a HWP which can rotate fast to scramble polarization at up to 20 Mrad/s. Since the orientation of the HWP rotations with respect to the PSPs of the DGD sections should be restricted as little as possible, one or better more waveplates (such as QWP) should be placed before and behind the HWP. One possible embodiment is shown in Fig. 3. It has the additional advantage that the relatively large total number of 7 waveplates makes it easier to achieve a Rayleigh-like speed distribution if this is desired.

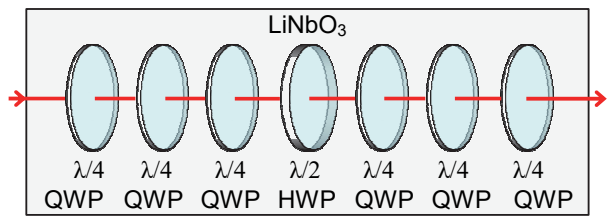

Fig. 3. Possible embodiment of a retarder/scrambler for a PMDE. Rotatable HWP can generate polarization transformations up to 20 Mrad/s. Rotatable QWPs before and behind permit arbitrary orientation of polarization transformations caused by HWP. All 7 waveplates are integrated in one packaged, commercially available LiNbO$_3$ chip.

The endless scrambling of the HWP should include definable bursts of forth/back rotation up to ±20 Mrad/s.

## V. TYPE AND DGD RANGE OF DGD SECTIONS

Each transponder or coherent RX manufacturer knows how much DGD the DSP can equalize. Therefore he can tell the tester how large the total DGD should be chosen. Each individual DGD can then be chosen as the total DGD divided by $N$. For instance, assuming 2.5 ps/m of DGD in PMF, a total DGD of 250 ps can be achieved with $N = 2$ DGD sections,

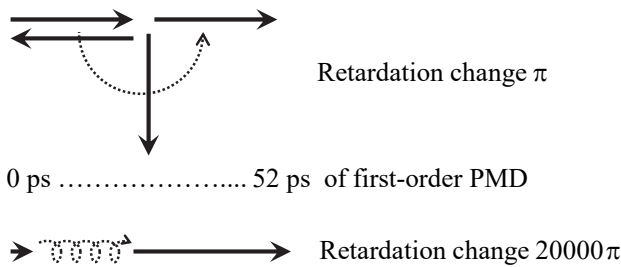

Fig. 4. Top: Transmission span containing 2 sections with 26 of DGD each most easily changes PMD by a mode converter (retarder) placed in between, capable of a π retardation change. Bottom: Pure DGD vector lengthening is extremely unlikely to happen in a transmission span.

each with a DGD of 125 ps and length of 50 m.

It has been detailed in [2] that an increase of total DGD is essentially not due to DGD sections changing their delay. If, for example, there are 2 sections with 26 ps of DGD each, a retardation change from π to 0 of a mode converter in between the sections change the total DGD from 0 ps to 52 ps (Fig. 4). In contrast, changing the DGD of a single DGD section from 0 ps to 52 ps needs about 20000·π, i.e. a 20000 times larger retardation change. Of course the smaller retardation change of π is many, many orders of magnitude more likely and will be decisive for all at least all non-minuscule changes of total DGD.

Since a certain range of total DGDs is to be covered by a very small number of DGD sections, maybe as few as $N = 2$ due to cost of the $N+1$ time-variable retarders/scramblers, one should consider changing the range of total covered DGD by changing the section DGDs, be it

- by PMF replacements,
- by optical switches, in particular optomechanical switches,

- by variable DGD sections.

Variable DGD elements are a versatile alternative to fixed DGD sections. If DSP tracking speed in the coherent receiver were slow, retardation changes of DGD in variable sections due thermal or wavelength drift alone could already be a challenge to the DSP. But given that DSP is very fast anyway, variable DGD sections can be tolerated.

A delay line alone is not a DGD section. Rather, 2 polarization beamsplitters/combiners are also needed. This increases cost somewhat and may introduce PDL.

It seems reasonable that it should be possible to achieve a neutral PMDE which no PMD or DGD whatsoever. A convenient implementation is it to choose an even $N$ with a symmetric distribution of section DGDs. This way the innermost sections $N/2$, $N/2+1$ can compensate each other, likewise the sections $N/2-1$, $N/2+2$, ... , ..., 1, $N$. We propose to make all section DGDs identical, with even $N$. Maximum total and first-order PMD/DGD is simply $N$ times section DGD.

In order to leave it open whether DGD sections are fixed or variable, a variety of total DGD values should be specified. Example: 20 ps, 50 ps, 100 ps, 200 ps. Finer granularity biases the engineering implementations toward usage of variable DGD sections, due to cost reasons.

One day it may become interesting to emulate not only distributed PMD but also distributed PDL. The proposed PMDE can easily be enhanced this way. See Section III. of [18] where a distributed DGD+PDL model is presented.

## VI. Availability of components and subsystems

LiNbO$_3$ polarization transformers with up to 8 waveplates are available from EOSPACE since many years [19] and they are recently also being manufactured by Fiberpro [20]. A shortage of such devices is therefore not to be expected, and competition will keep component prices reasonable.

Manufacturers of polarization scramblers or time-variable retarders with such LiNbO$_3$ polarization transformers are seemingly (in lexical order) Keysight [21], New Ridge Technologies [22], Novoptel [23] and Viavi [24]. Maximum scrambling speed 20 Mrad/s [23] or even 50 Mrad/s [16] is available from Novoptel. As far as not all manufacturers currently offer the proposed maximum scrambling speed of 20 Mrad/s and have a satisfying number of waveplates (7 in [23]) it is to be assumed that they can implement adequate changes to achieve this, given that the same LiNbO$_3$ polarization transformers are available for everyone.

PMF for fixed DGD sections and optomechanical switches to switch between DGD sections of different lengths are available from a number of vendors at competitive prices. Any vendor of polarization scramblers can integrate them into his boxes.

Such a complete PMDE with 20 Mrad/s polarization scramblers and switchable DGD sections (internal, but also external ones) is available from Novoptel [25].

A motorized fiberoptic DGD section with a range of 0...200 ps is available from Luna (General Photonics) [26].

New Ridge Technologies also offers a PMDE [27], and so does Luna (General Photonics) [28]. It is unclear whether these are constructed of DGD sections as needed for physical modeling.

From the above it is clear that existing competition can always guarantee availability of the proposed PMDE at reasonable cost.

## VII. Summary of recommendations for PMDE in long-range, high-end coherent transmission

For the implementation of high-end PMDE we recommend:

- $N$ DGD sections placed between $N+1$ time-variable retarders/scramblers should be used to emulate PMD. The larger $N$, the better, but $N = 2$ is already an acceptable choice.
- Even $N$ with equal section DGDs is recommended because it includes the possibility of having a neutral PMDE.
- A variety of total DGD (= 1st order PMD) values should be specified, say 20 ps, 50 ps, 100 ps, 200 ps, in order to keep the choice between variable and fixed DGD sections open.
- The scramblers should be elliptical retarders, capable of endless polarization transformations with speeds up to 20 Mrad/s, including definable forth/back rotations up to that speed with arbitrarily adjustable orientation.

The PMDEs can be supplied as complete units, or in parts, namely time-variable retarders/scramblers and DGD sections, from different or identical vendors.

## VIII. Recommendations for PMDE in coherent 80 km ZR link

In a coherent 80 km ZR link [29] the boundary conditions are 10 ps mean DGD and 50 krad/s maximum scrambling speed. For the implementation of a PMDE we recommend:

- 2 DGD sections placed between 3 time-variable retarders/scramblers seem acceptable.
- DGD per section equals total maximum DGD (e.g. 3 times mean DGD, hence 30 ps) divided by $N$ (hence 15 ps).
- The scramblers should be elliptical retarders, capable of endless polarization transformations with speeds up to 50 krad/s.

For those testers who also expect to need a PMDE for long-range, high-end coherent transmission it may prove most cost-effective to choose a model that fulfills, or can be modified to fulfill, the higher requirements of Section VII.

## IX. Acknowledgement

We thank Dr. Bo Zhang (Inphi Corp.) for suggesting the writing of this text and for very helpful discussions on this topic.